\documentclass[10pt,a4paper,superscriptaddress,showpacs,notitlepage,twocolumn]{revtex4}

\usepackage{amssymb}
\usepackage{amsfonts}
\usepackage{hyperref}
\usepackage[dvips]{graphicx}
\usepackage{wrapfig}

\begin{document}

\title{Truncated states obtained by iteration}
\author{W. B. Cardoso}
\email[Corresponding author: ]{wesleybcardoso@gmail.com}
\affiliation{Instituto de F\'{\i}sica, Universidade Federal de Goi\'{a}s, 74.001-970, Goi%
\^{a}nia (GO), Brazil}
\author{N. G. de Almeida}
\affiliation{N\'{u}cleo de Pesquisas em F\'{\i}sica, Universidade Cat\'{o}lica de Goi\'{a}%
s, 74.605-220, Goi\^{a}nia (GO), Brazil.}
\affiliation{Instituto de F\'{\i}sica, Universidade Federal de Goi\'{a}s, 74.001-970, Goi%
\^{a}nia (GO), Brazil}

\begin{abstract}
Quantum states of the electromagnetic field are of considerable importance,
finding potential application in various areas of physics, as diverse as
solid state physics, quantum communication and cosmology. In this paper we
introduce the concept of truncated states obtained \textit{via} iterative
processes (TSI) and study its statistical features, making an analogy with
dynamical systems theory (DST). As a specific example, we have studied TSI
for the doubling and the logistic functions, which are standard functions in
studying chaos. TSI for both the doubling and logistic functions exhibit
certain similar patterns when their statistical features are compared from
the point of view of DST. A general method to engineer TSI in the
running-wave domain is employed, which includes the errors due to the
nonidealities of detectors and photocounts.
\end{abstract}


\maketitle

\section{Introduction}

Quantum state engineering is an area of growing importance in quantum
optics, its relevance lying mainly in the potential applications in other
areas of physics, such as quantum teleportation \cite{bennett93}, quantum
computation \cite{Kane98}, quantum communication \cite{Pellizzari97},
quantum cryptography \cite{Gisin02}, quantum lithography \cite{Bjork01},
decoherence of states \cite{Zurek91}, and so on. To give a few examples of
their usefulness and relevance, quantum states arise in the study of quantum
decoherence effects in mesoscopic fields \cite{harochegato}; entangled
states and quantum correlations \cite{brune}; interference in phase space
\cite{bennett2}; collapses and revivals of atomic inversion \cite{narozhny};
engineering of (quantum state) reservoirs \cite{zoller}; etc. Also, it is
worth mentioning the importance of the statistical properties of one state
in determining some relevant properties of another \cite{barnett}, as well
as the use of specific quantum states as input to engineer a desired state
\cite{serra}.

Dynamical Systems Theory (DST), on the other hand, is a completely different
area of study, whose interest lies mainly in nonlinear phenomena, the source
of chaotic phenomena. DST groups several approaches to the study of chaos,
involving Lyapunov exponent, fractal dimension, bifurcation, and symbolic
dynamics among other elements \cite{devaney}. Recently, other approaches
have been considered, such as information dynamics and entropic chaos degree
\cite{ohya}.

The purpose of this paper is twofold: to introduce novel states of
electromagnetic fields, namely truncated states having coefficients obtained
\textit{via} iterative process (TSI), and to study chaos phenomena using
standard techniques from quantum optics, making an analogy with DST. We note
that, unlike previous states studied in the literature \cite{dodonov}, each
coefficient of the TSI is obtained from the previous one by iteration of a
function. Features of this state are studied by analyzing several of its
statistical properties in different regimes (chaotic \textit{versus}
nonchaotic) according to DST, and, for some iterating functions, we found
properties of TSI very sensitive (resembling chaos) to the first coefficient
$C_{0}$, which is used as a ~ seed ~ to obtain the remaining $C_{n}$.

This paper is organized as follows. In section 2 we introduce the TSI and in
section 3 we analyze the behavior of some of its properties as the Hilbert
space dimension is increased. In section 4 we show how to engineer the TSI
in the running-wave field domain, and the corresponding engineering fidelity
is studied in section 5. In section 6 we present our conclusions.

\section{Truncated states obtained via iteration (TSI)}

We define TSI as
\begin{equation}
|TSI\rangle =\sum\limits_{n=0}^{N}C_{n}|n\rangle
\end{equation}%
where $C_{n}$ is the normalized complex coefficient obtained as the $n$th
iteration of a previously given generating function. For example, given $%
C_{0}$,$C_{n}$ can be the $n$th iterate of the quadratic functions: $%
C_{n}(\mu )=C_{n-1}^{2}+\mu $; sine functions: $C_{n}(\mu )=\mu \sin
(C_{n-1})$; logistic functions $C_{n}(\mu )=\mu C_{n-1}(1-C_{n-1})$;
exponential functions: $C_{n}=\mu \exp (C_{n-1})$; doubling function defined
on the interval [0,1): $C_{n}=2C_{n-1}$ \textit{mod} $1$, and so on, $\mu $
being a parameter. It is worth recalling that all the functions in the above
list are familiar to researchers in the field of dynamical systems theory
(DST). For example, for some values of $\mu $, it is known that some of
these functions can behave in quite a chaotic manner \cite{devaney}. Also,
note that by computing all the $C_{n}$ we are in fact determining the
\textit{orbit} of a given function, and because the $C_{n}$ and $P_{n}$, the
photon number distribution, are related by $P_{n}=|C_{n}|^{2}$, fixed or
periodic points of a function will correspond to fixed or periodic $P_{n}$.
Rather than studying all the functions listed in this section, we will focus
on the doubling function and the logistic function. These two functions have
been widely used to understand chaos in nature. As we shall see in the
following, although very different from each other, these functions give
rise to different TSI having similar patterns.

\section{Statistical properties of TSI using the doubling and the logistic
functions}

\subsection{Photon Number Distribution}

Since the expansion of TSI is known in the number state $|n\rangle $, we
have
\begin{equation}
P_{n}=|C_{n}|^{2}.  \label{pn}
\end{equation}%
Figs. $1$ and $2$ show the plots of the photon-number distribution $P_{n}$
versus $n$ for TSI using the doubling function. The Hilbert space dimension
is $N=50$. In order to illuminate the behavior of TSI for different values
of $C_{0}$, we take $C_{0}$ as $0.3$ and $0.29711$, respectively shown in
Figs. $1$ and $2$. Note the regular behavior for $C_{0}=0.3$ and rather an
irregular, or chaotic, behavior for $C_{0}=$ $0.29711$. Figs. $3$ and $4$
show $P_{n}$ for the logistic function. For $C_{0}=0.2$ and $\mu =3.49$ the
logistic function behaves regularly (Fig.$3$), showing clearly (as in the
case of the doubling-function) four values for $P_{n}$; by contrast, for $%
\mu =4$ and $C_{0}=0.2$, $P_{n}$ oscillates quite irregularly (Fig.$4$).
This is so because the photon number distribution is equivalent to the
\textit{orbit} of the TSI dynamics \cite{devaney}. Thus, once a fixed -
attracting or periodic - point is attained, the subsequent coefficients, and
hence the subsequent $P_{n}$, will behave in a regular manner. Conversely,
when no fixed point exists, $P_{n}$ will oscillate in a chaotic manner.
Therefore, by choosing suitable $C_{0}$ and/or $\mu $, we can compare the
properties of TSI when different regimes (chaotic \textit{versus}
nonchaotic) in the DST sense are encountered. Note the similarity between
the properties of the logistic and the doubling functions when the DST
regimes are the same. Interestingly, these similarities are observed when
other properties are analyzed, as we shall see in the following.

\begin{figure}[t!]
\includegraphics[width=7cm, height=7cm]{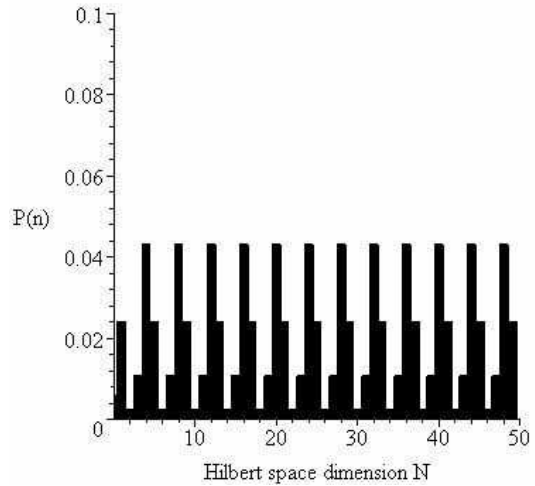}
\caption{Photon number distribution for the doubling function with $%
C_{0}=0.3 $. Note a four-period type for the probabilities; this regular
behavior coincides with nonchaotic behavior of the doubling function in the
DST sense. }
\end{figure}

\begin{figure}[t!]
\includegraphics[width=7cm, height=7cm]{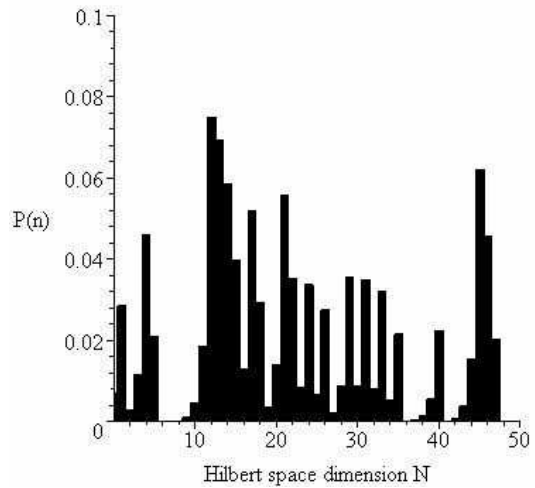}
\caption{Photon number distribution for the doubling function with $%
C_{0}=0.29711$. This irregular behavior for $P_{n}$ coincides with the
chaotic behavior of the doubling function in the DST sense.}
\end{figure}

\begin{figure}[t]
\includegraphics[width=7cm, height=7cm]{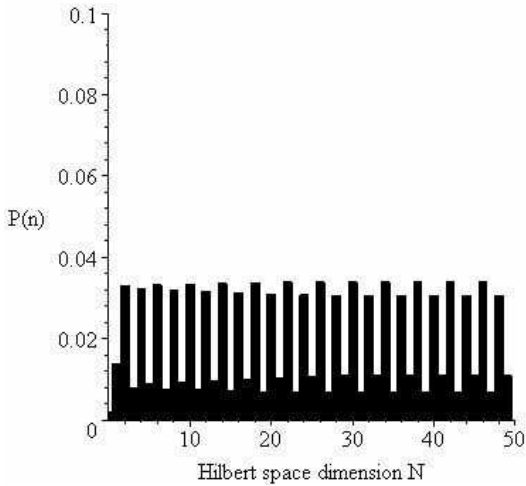}
\caption{Photon number distribution for the logistic function with $%
C_{0}=0.2 $ and $\protect\mu =3.49$. This regular or \textquotedblleft
four-period\textquotedblright\ type behavior for the probabilities coincides
with nonchaotic behavior in the DST sense.}
\end{figure}

\begin{figure}[t]
\includegraphics[width=7cm, height=7cm]{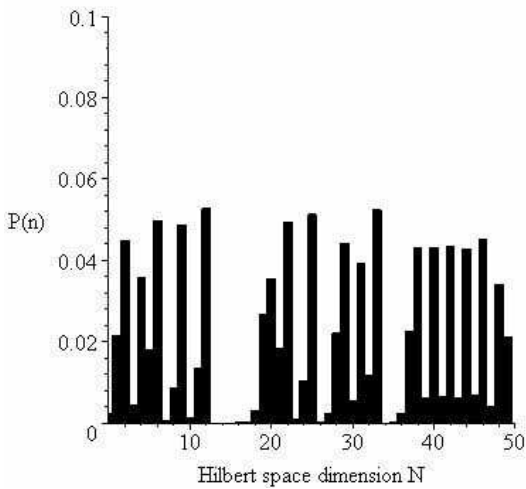}
\caption{Photon number distribution for the logistic function with $%
C_{0}=0.2 $ and $\protect\mu =4$. This irregular behavior of $P_{n}$
coincides with the chaotic behavior of the logistic function in the DST
sense.}
\end{figure}

\subsection{Even and Odd Photon number distribution}

The functions $P_{odd}$ and $P_{even}$ represent the photon number
distribution for $n$ $odd$ and $even$, respectively, given by Eq. (\ref{pn}%
). It is well established in quantum optics \cite{Mandel} that if $%
P_{odd}>0.5$ the Glauber-Sudarshan $P$-function assumes negative values,
prohibited in the usual probability distribution function, and the quantum
state has no classical analog. Since $P_{odd}+P_{even}=1$, the same is true
when $P_{even}<0.5$ . Figs. $5$ and $6$ show the behavior of $P_{odd}$ for $%
C_{0}=0.3$ and $C_{0}=0.29711$ for the doubling function the Hilbert space $N
$ is increased. Figs. $7$ and $8$ refer to the the logistic function for $%
\mu =3.49$ and $\mu =4$. In Figs. $5$ and $7$ (corresponding to a nonchaotic
regime in DST), note that TSI has a classical analog as $N$ increases. From
Figs. $6$ and $8$ (corresponding to a chaotic regime in DST), TSI can behave
as a nonclassical state, depending on $N$. More interestingly, note the
following pattern: whenever the coefficients of TSI correspond to the
nonchaotic regime in DST, $P_{odd}$ (and so $P_{even}$) will remain above or
below $0.5$ on a nearly monotonic curve, as seen in Figs. $5$ and $7$;
whenever the coefficients of TSI correspond to the chaotic regime in DST, $%
P_{odd}$ (and so $P_{even}$) will tend to oscillate around $0.5$ (Figs. $6$
and $8$).

\begin{figure}[tb]
\includegraphics[width=8cm, height=6.2cm]{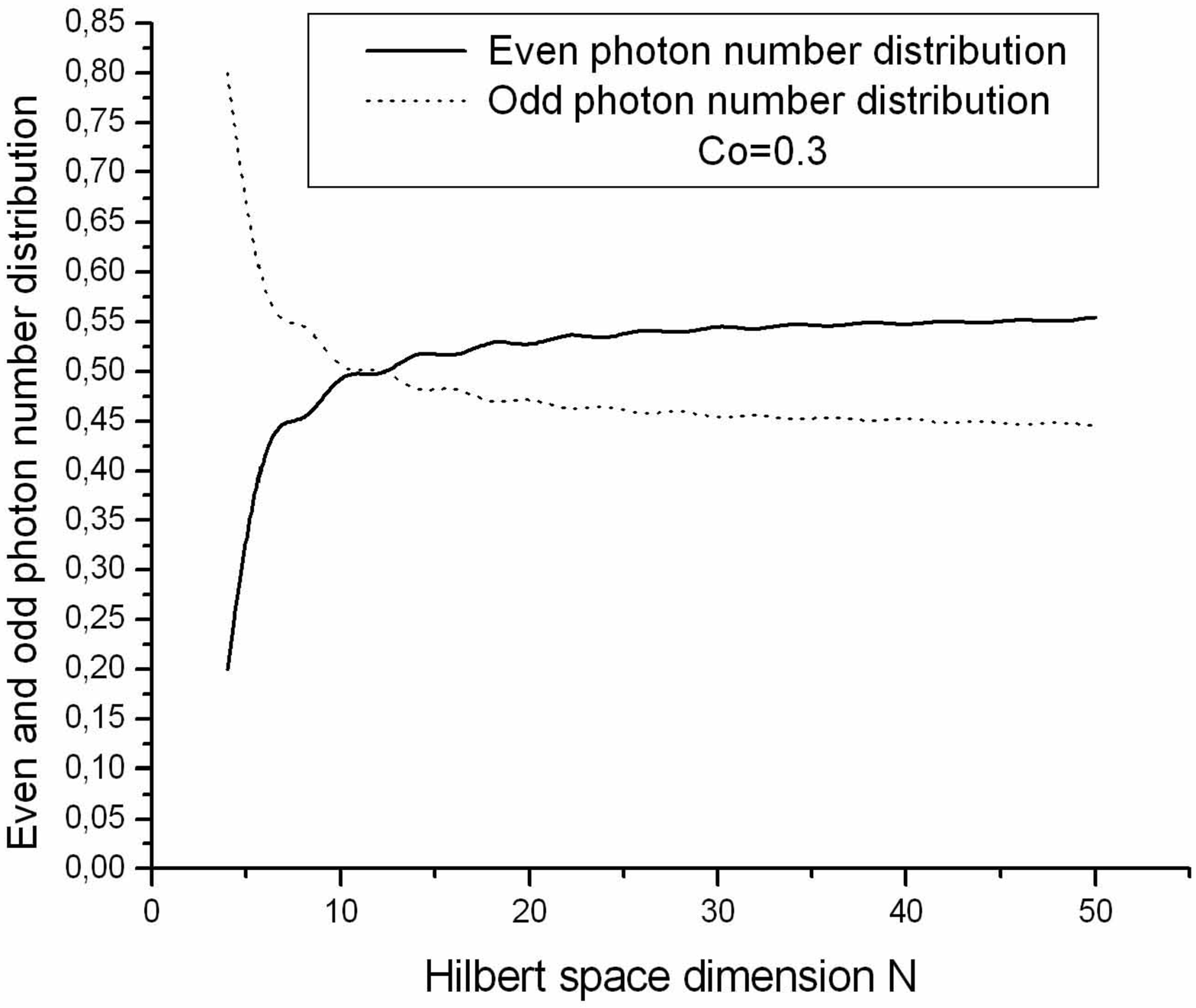}
\caption{Even (solid) and odd (dots) photon number distributions for the
doubling function. We have used $C_{0}=0.3$ to coincide with nonchaotic
behavior in the DST sense.}
\end{figure}

\begin{figure}[tb]
\includegraphics[width=8cm, height=6.2cm]{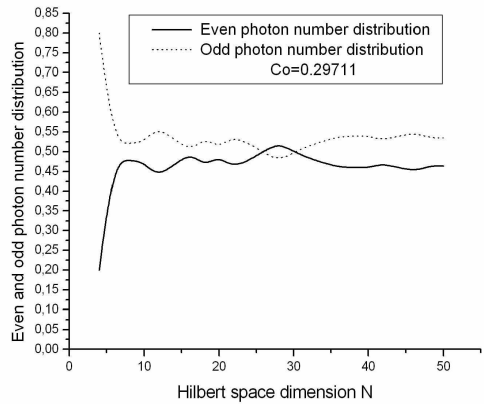}
\caption{Even (solid) and odd (dots) photon number distributions for the
doubling function. $C_{0}=0.29711$ was chosen to coincide with chaotic
behavior in the DST sense.}
\end{figure}

\begin{figure}[tb]
\includegraphics[width=8cm, height=6.2cm]{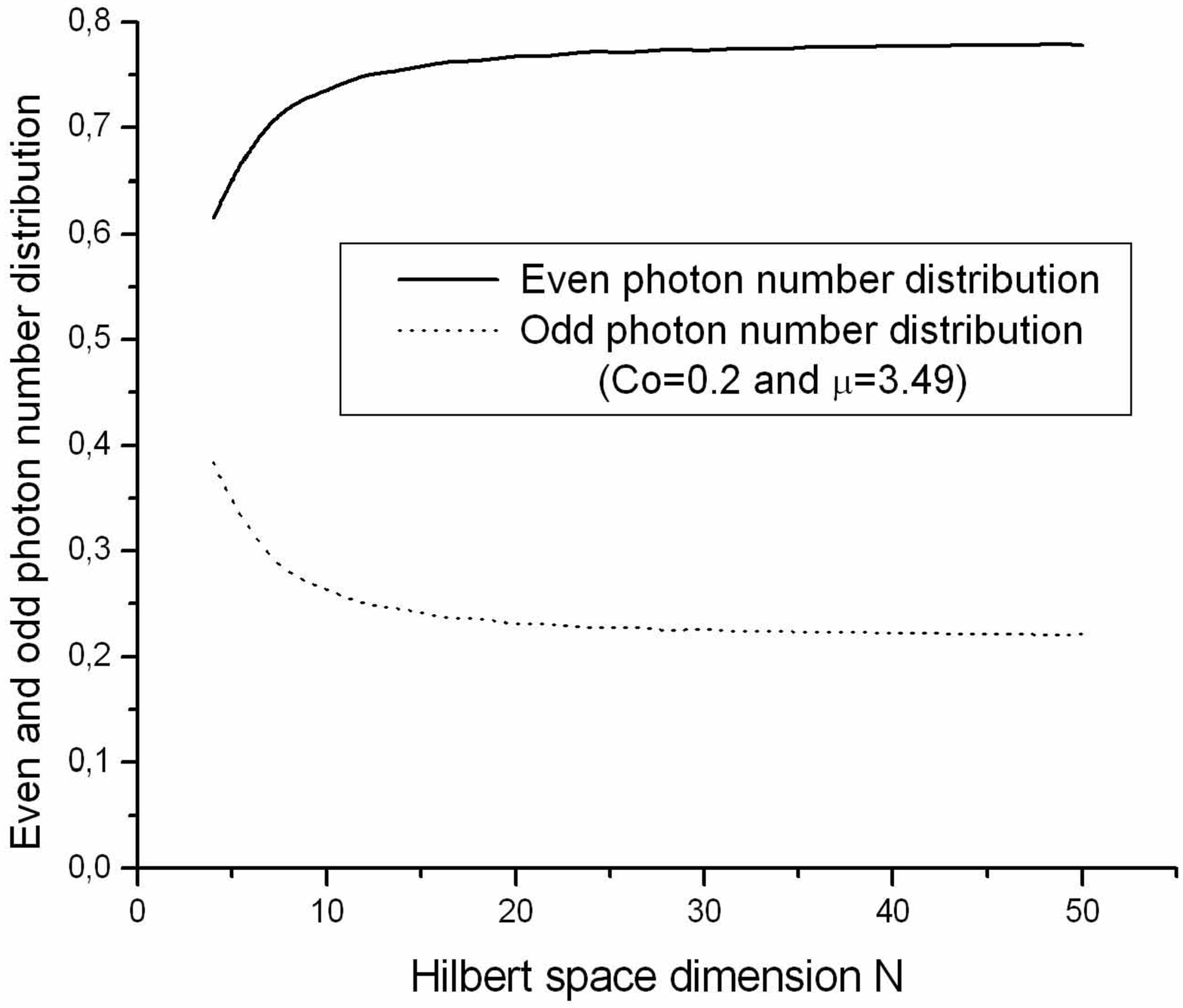}
\caption{Even (solid) and odd (dots) photon number distributions for the
logistic function. Here we chose $C_{0}=0.2$ and $\protect\mu =3.49$ to
coincide with nonchaotic behavior in the DST sense.}
\end{figure}

\begin{figure}[tb]
\includegraphics[width=8cm, height=6.2cm]{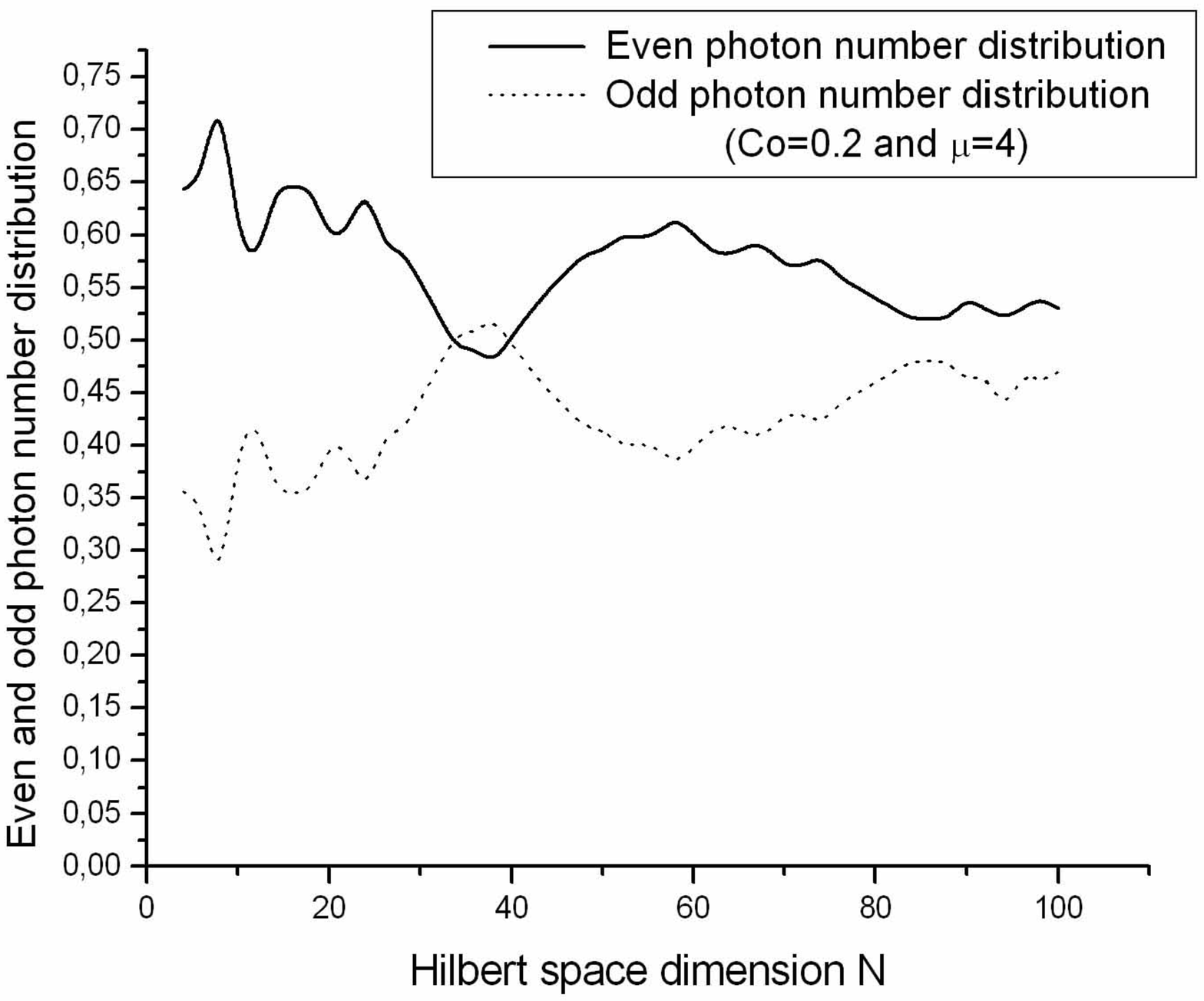}
\caption{Even (solid) and odd (dots) photon number distributions for the
logistic function. Here we chose $C_{0}=0.2$ and $\protect\mu =4$ to
coincide with chaotic behavior in the DST sense.}
\end{figure}

\subsection{Average number and variance}

The average number $\left\langle \hat{n}\right\rangle $ and the variance $%
\left\langle \Delta\hat{n}\right\rangle $ in TSI are obtained
straightforwardly from

\begin{equation}
\left\langle \hat{n}\right\rangle =\sum_{n=0}^{N}P(n)n ,
\end{equation}
and

\[
\left\langle \Delta\hat{n}\right\rangle =\sqrt{\left\langle \hat{n}%
^{2}\right\rangle -\left\langle \hat{n}\right\rangle ^{2}} .
\]

Fig. $9$ shows the plot of $\left\langle \hat{n}\right\rangle $ and Fig. $10$
the plot of $\left\langle \Delta\hat{n}\right\rangle $ as functions of the
dimension $N$ of Hilbert space, for the doubling function. Note the near
linear behavior of the average photon number and its variance as $N$
increases for $C_{0}=0.3$ (nonchaotic regime in DST); this is not seen when $%
C_{0}=0.29711$ (chaotic regime in DST). Figs. $11$ and $12$ for the logistic
function show essentially the same behavior when these two DST regimes are
shown together.

\begin{figure}[tb]
\includegraphics[width=8cm, height=6.2cm]{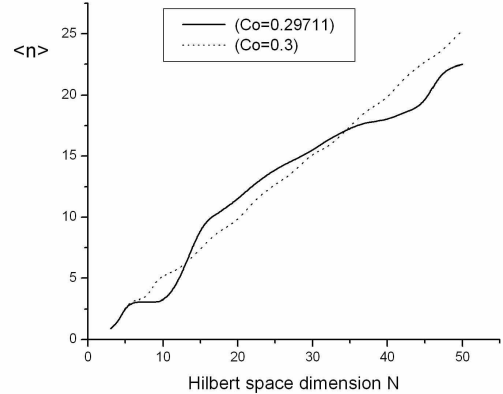}
\caption{Average photon number for the doubling function. Here we chose $%
C_{0}=0.3$ (dots) and $C_{0}=0.29711$ (solid), allowing the Hilbert space $N$
to increase to $50$.}
\end{figure}

\begin{figure}[tb]
\includegraphics[width=8cm, height=6.2cm]{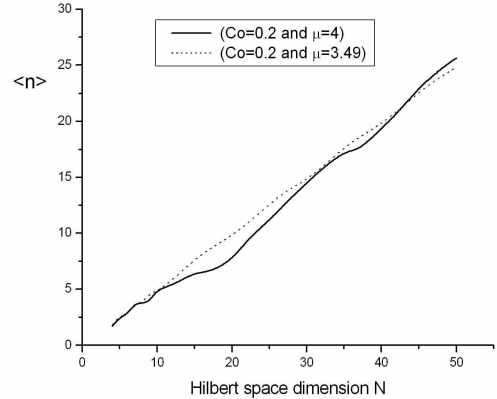}
\caption{Variance of the photon number for the doubling function. Here we
chose $C_{0}=0.3$ (dots)\ as well as $C_{0}=0.29711$(solid), allowing the
Hilbert space $N$ to increase to $50$.}
\end{figure}

\begin{figure}[tb]
\includegraphics[width=8cm, height=6.2cm]{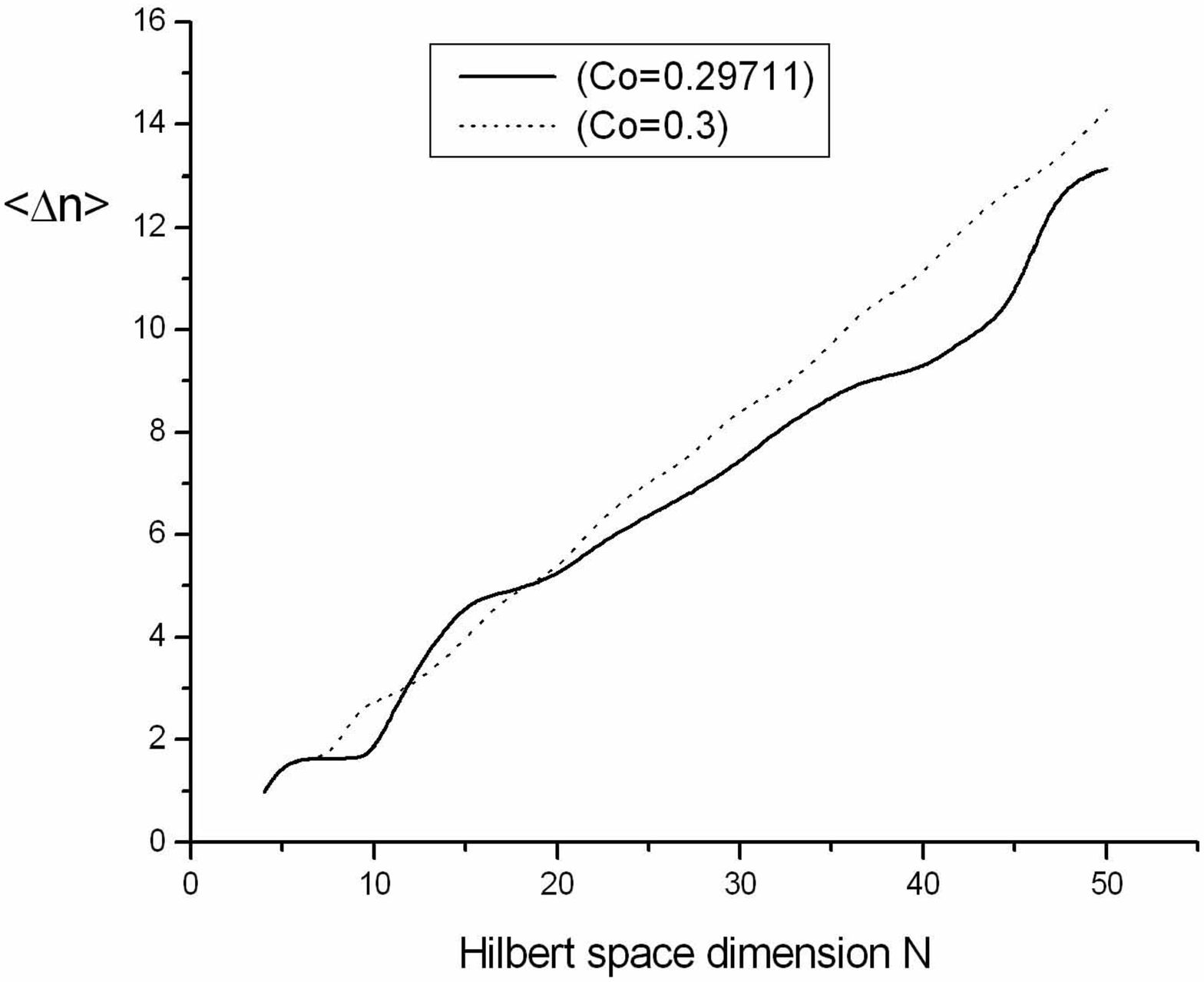}
\caption{Average photon number for the logistic function. Here we chose $%
C_{0}=0.2$ with $\protect\mu =3.49$ (dots) or $\protect\mu =4$ (solid),
allowing the Hilbert space $N$ to increase to $50$.}
\end{figure}

\begin{figure}[tb]
\includegraphics[width=8cm, height=6.2cm]{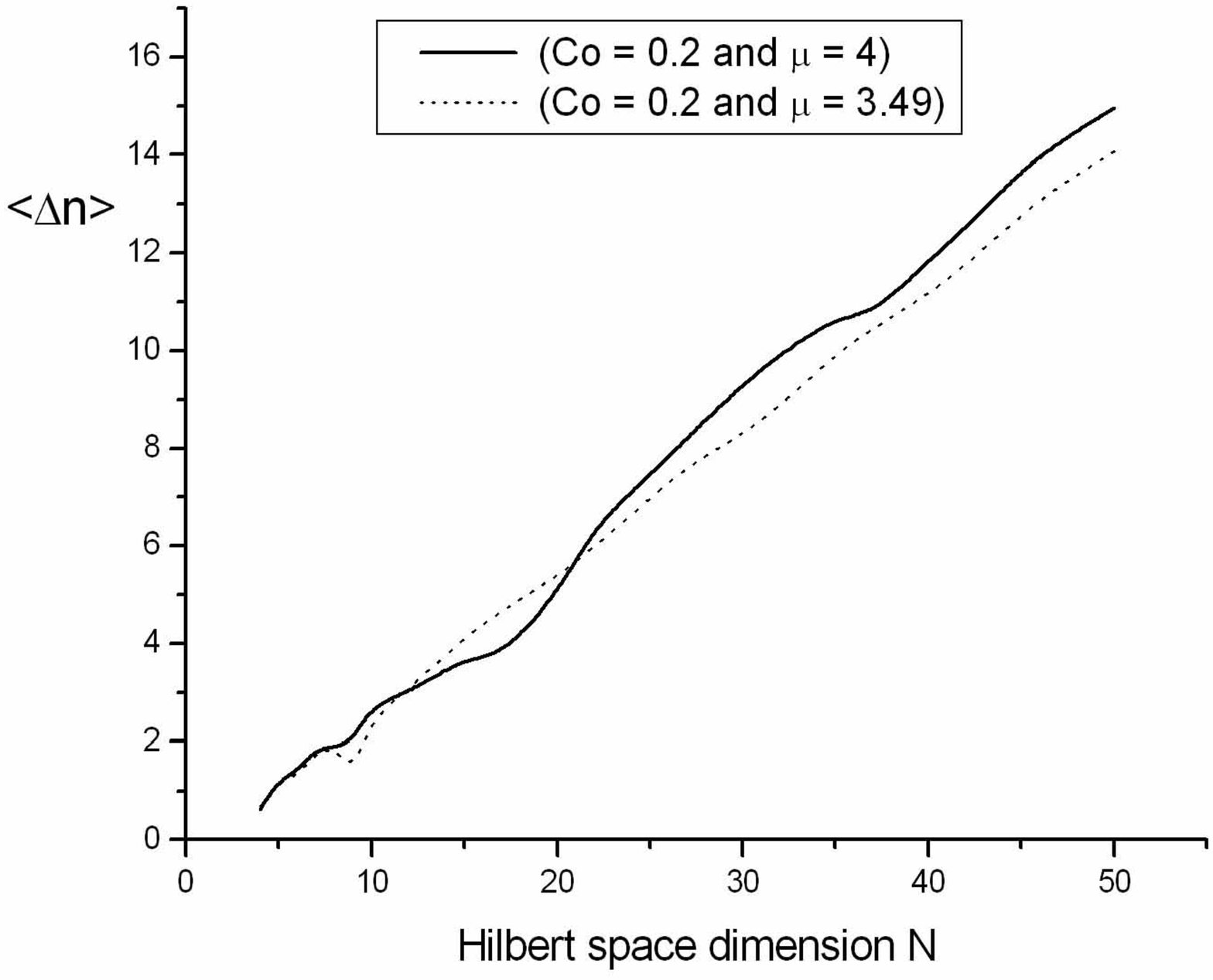}
\caption{Variance of the photon number for the logistic function. Here we
chose $C_{0}=0.2$ with $\protect\mu =3.49$ (dots) or $\protect\mu =4$
(solid), allowing the Hilbert space $N$ to increase to $50$.}
\end{figure}

\subsection{Mandel parameter and second order correlation function}

The Mandel $Q$ parameter is defined as

\begin{equation}
Q=\frac{\left( \Delta\hat{n}^{2}-\left\langle \hat{n}\right\rangle \right) }{%
\left\langle \hat{n}\right\rangle } ,
\end{equation}
while the second order correlation function $g^{\left( 2\right) }(0)$ is
\[
g^{\left( 2\right) }(0)=\frac{\left( \left\langle \hat{n}^{2}\right\rangle
-\left\langle \hat{n}\right\rangle \right) }{\left\langle \hat{n}%
\right\rangle ^{2}} ,
\]
and for $Q<0$ ($Q>0$) the state is said to be sub-Poissonian
(super-Poissonian). Also, the $Q$ parameter and the second order correlation
function $g^{\left( 2\right) }$ are related by \cite{walls}
\begin{equation}
Q=\left[ g^{\left( 2\right) }(0)-1\right] \left\langle \hat{n}\right\rangle .
\label{qg}
\end{equation}

If $g^{\left( 2\right) }(0)<0$,\ then the Glauber-Sudarshan $P$-function
assumes negative values, outside the range of the usual probability
distribution function. Moreover, by Eq. (\ref{qg}) it is readily seen that $%
g^{\left( 2\right) }(0)<1$ implies $Q<0$. As for a coherent state $Q=0$, a
given state is said to be a ``classical''\ one if $Q>0$.

Figs. $13$ to $16$ show the plots of the $Q$ parameter and the correlation
function $g^{\left( 2\right) }(0)$ versus $N$ ,\ for both the doubling and
the logistic functions. Note that TSI is predominantly super-Poissonian for
these two functions ($Q>0$ and $g^{\left( 2\right) }(0\dot{)}>1$), thus
being a \textquotedblleft classical\textquotedblright\ state in this sense
for $N\gtrsim 12$, while for small values of \ $N$ ($N<12$), the $Q$
parameter is less than $0$, showing sub-Poissonian statistics and is thus
associated with a \textquotedblleft quantum state\textquotedblright . From
Figs.$13$ and $15$, note that using $C_{0}=0.3$ for the doubling function, $%
C_{0}=0.2$ and $\mu =3.49$ for the logistic function (nonchaotic regime in
DST), $Q$ shows a linear dependence on $N$. However, using $C_{0}=0.29711$
for the doubling function, $C_{0}=0.2$ and $\mu =4$ for the logistic
function (chaotic regime in DST), $Q$ oscillates irregularly. Similarly,
from Figs. $14$ and $16$, using $C_{0}=0.3$ for the doubling function, $%
C_{0}=0.2$ and $\mu =3.49$ for the logistic function, we see that $g^{\left(
2\right) }(0)$ increases smoothly, while using $C_{0}=0.29711$ for the
doubling function, $C_{0}=0.2$ and $\mu =4$ for the logistic function, the
rise of $g^{\left( 2\right) }(0)$ is rather irregular. This pattern is
observed for other $C_{0}$ as input as well as for other values of the
parameter $\mu $, and whenever the dynamics is chaotic (regular), the $Q$
parameter and the $g^{\left( 2\right) }(0)$ correlation function oscillate
irregularly (regularly).

\begin{figure}[tb]
\includegraphics[width=8cm, height=6.2cm]{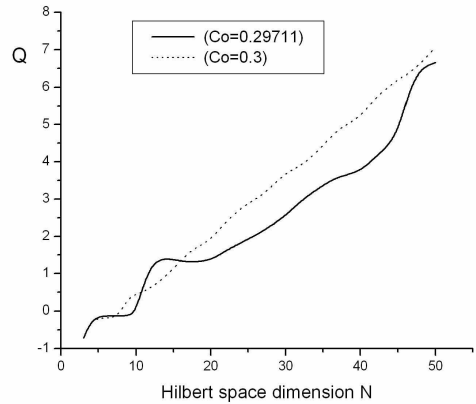}
\caption{Q parameter for the doubling function. Here we chose $C_{0}=0.3$
(dots) \ and $C_{0}=0.29711$ (solid), allowing the Hilbert space $N$ to
increase to $50$. Note the irregular behavior of the Q parameter when $%
C_{0}=0.29711$, coinciding with chaotic behavior in the DST sense.}
\end{figure}

\begin{figure}[tb]
\includegraphics[width=8cm, height=6.2cm]{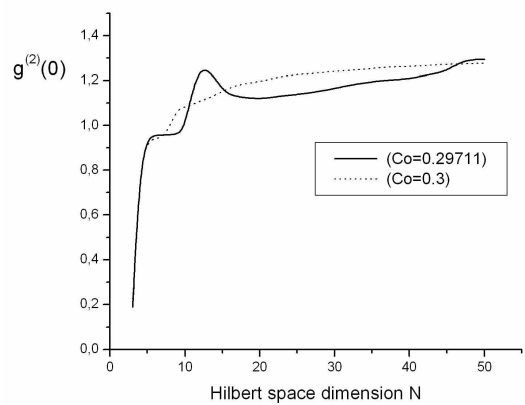}
\caption{Second order correlation function for the doubling function. Here
we chose $C_{0}=0.3$ (dots) and $C_{0}=0.29711$ (solid), allowing the
Hilbert space $N$ to increase to $50$. Note the irregular rise of the $g(0)$
function when $C_{0}=0.29711$, coinciding with chaotic behavior in the DST
sense.}
\end{figure}

\begin{figure}[tb]
\includegraphics[width=8cm, height=6.2cm]{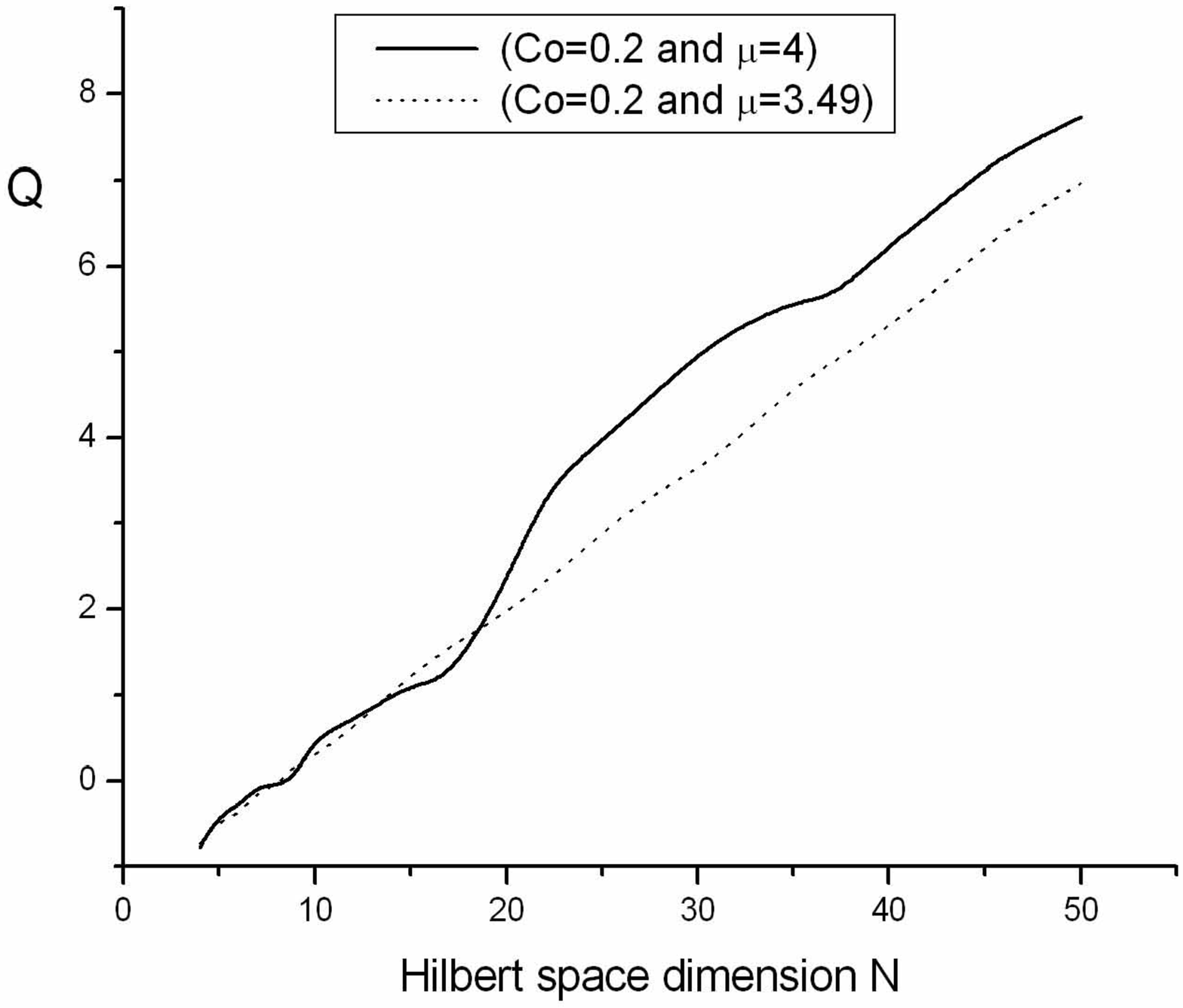}
\caption{Q parameter for the logistic function. Here we chose $C_{0}=0.2$,
with $\protect\mu =3.49$ (dots) and $\protect\mu =4$ (solid), allowing the
Hilbert space $N$ to increase up to $50$. Note the irregular behavior of the
Q parameter when $\protect\mu =4$, exactly when chaotic behavior occurs
according to DST.}
\end{figure}

\begin{figure}[tb]
\includegraphics[width=8cm, height=6.2cm]{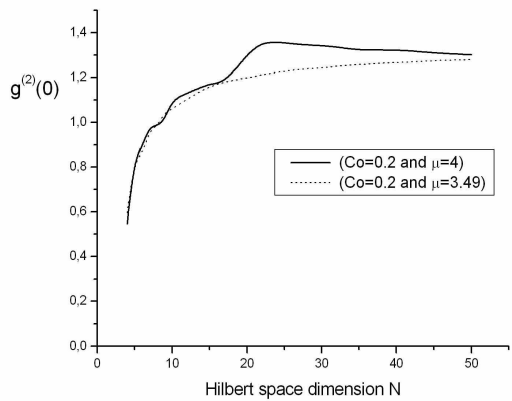}
\caption{Second order correlation function for the logistic function. Here
we chose $C_{0}=0.2$, with $\protect\mu =3.49 $\ (dots) and $\protect\mu =4$
(solid), allowing the Hilbert space $N$ to increase up to $50$. Note the
irregular rise of $g^{2}(0)$ when $\protect\mu =4$, exactly when chaotic
behavior occurs according to DST.}
\end{figure}

\subsection{Quadrature and variance}

Quadrature operators are defined as
\begin{eqnarray}
X_{1}& =\frac{1}{2}\left( a+a^{\dagger }\right) ;  \nonumber \\
X_{2}& =\frac{1}{2i}\left( a-a^{\dagger }\right) .
\end{eqnarray}%
where $a$ ($a^{\dagger }$) \ is the annihilation (creation) operator in Fock
space. Quantum effects arise when the variance of one of the two quadratures
attains a value $\Delta X_{i}<0.5$, $i=1,2$. Figs. $17$ and $18$ show the
plots of quadrature variance $\Delta X_{i}$ versus $N$. Note in this figures
that variances increase when $N$ is increased.

\begin{figure}[tb]
\includegraphics[width=8cm, height=6.2cm]{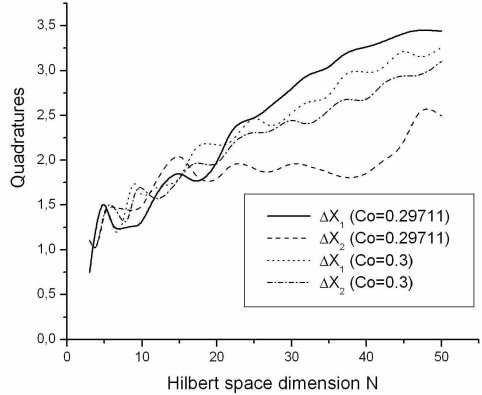}
\caption{Averaged quadratures for the doubling function when the Hilbert
space $N$ is increased to $50$. Doted (dashed-doted) line refer to
quadrature variance $1$ $(2)$ for $C_{0}=0.3$; solid (dashed) line refer to
quadrature variance 1 $(2)$ for $C_{0}=0.29711$.}
\end{figure}

\begin{figure}[tb]
\includegraphics[width=8cm, height=6.2cm]{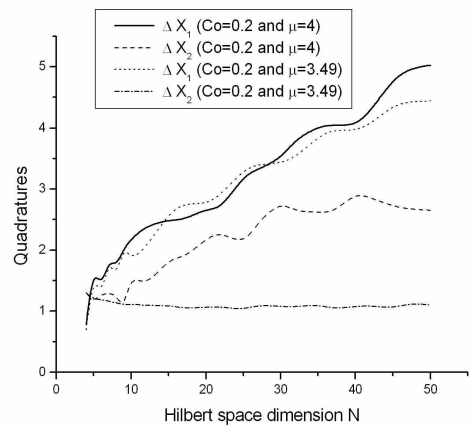}
\caption{Averaged quadratures for the logistic function with $C_{0}=0.2$
when the Hilbert space N is increased to $50$. Doted (dashed-doted) line
refer to quadrature variance $1$ ($2$) for $\protect\mu =3.49$; solid
(dashed) refer to quadrature variance 1 ($2$) for $\protect\mu =4$.}
\end{figure}

\subsection{Husimi -Q function}

The Husimi Q-function for TSI is given by
\begin{equation}
Q_{\left\vert TSI\right\rangle }(\beta )=\frac{1}{\pi }\left\vert
\left\langle \beta |TSI\right\rangle \right\vert ^{2},
\end{equation}%
where $\beta $ is a coherent state. Figs. $19$ to $22$ show the Husimi
Q-function for both the doubling and the logistic functions for $N=15.$ For
the doubling function, we use $C_{0}=0.29711$ and $C_{0}=0.3$, respectively,
and for the logistic function we use $\mu =3.49$ and $\mu =4$, respectively,
for the chaotic and nonchaotic regimes. Interestingly, even when the chaotic
and nonchaotic regimes of the DST are compared, Husimi Q-functions show
essentially no difference from each other.

\begin{figure}[tb]
\includegraphics[width=3.0009in, height=3.0009in]{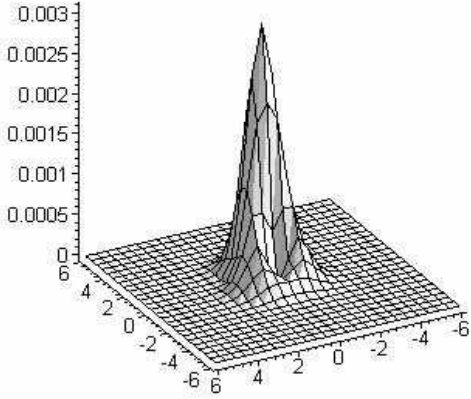}
\caption{Husimi Q function for doubling function. Here $C_{0}=0.3$.}
\end{figure}

\begin{figure}[tb]
\includegraphics[width=3.0009in, height=3.0009in]{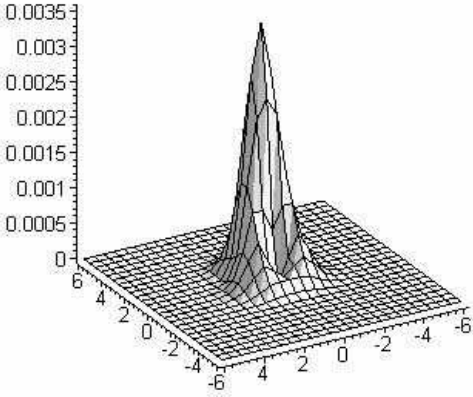}
\caption{Husimi Q function for doubling function. Here $C_{0}=0.29711$.}
\end{figure}

\begin{figure}[tb]
\includegraphics[width=3.0009in, height=3.0009in]{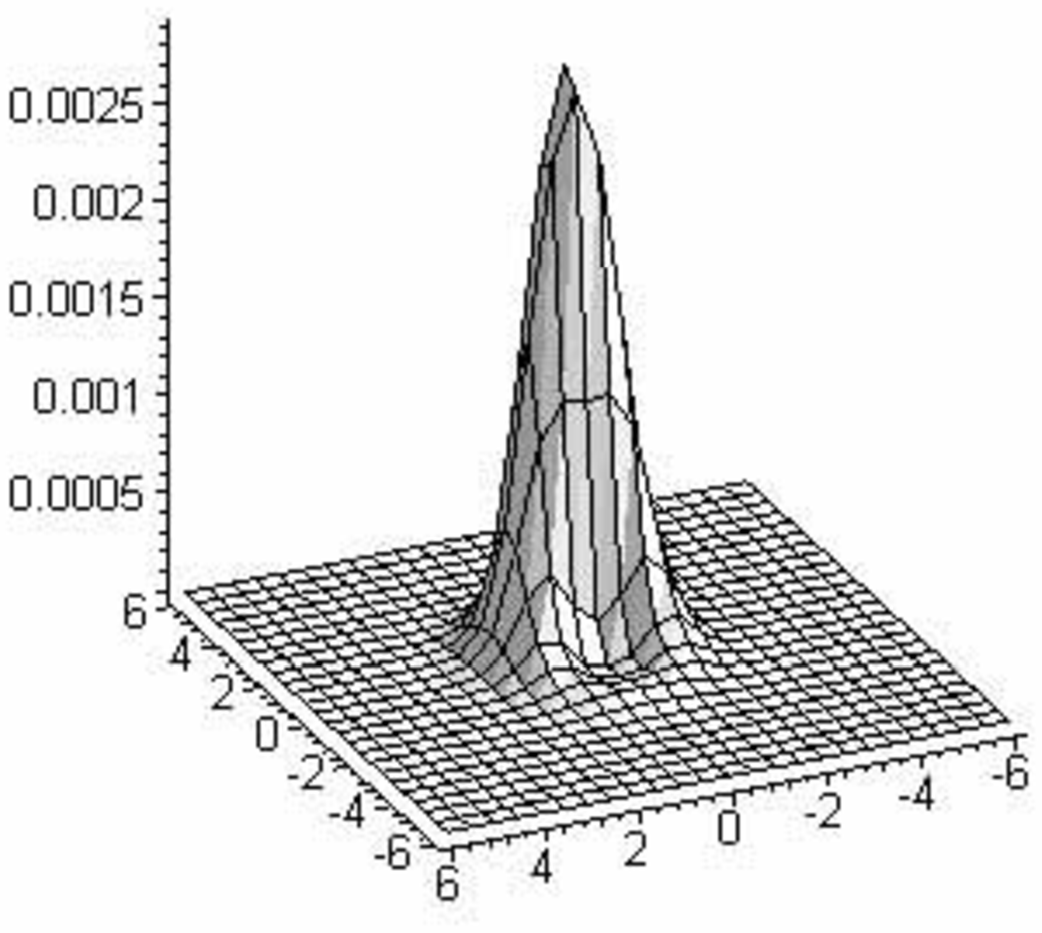}
\caption{Husimi Q function for logistic function. Here $C_{0}=0.2$ and $%
\protect\mu =3.49$.}
\end{figure}

\begin{figure}[tb]
\includegraphics[width=3.0009in, height=3.0009in]{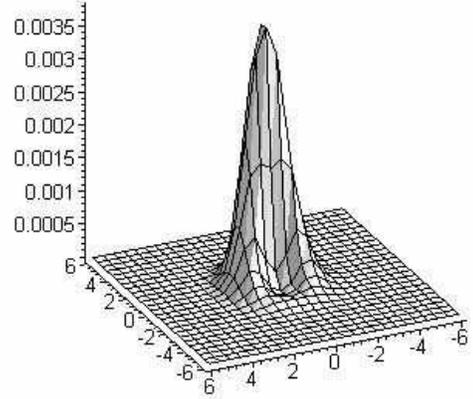}
\caption{Husimi Q function for logistic function. Here $C_{0}=0.2$ and $%
\protect\mu =4$.}
\end{figure}

\section{Generation of TSI}

TSI can be generated in various contexts, as for example trapped ions \cite%
{serra2}, cavity QED \cite{serra,vogel}, and travelling wave-fields \cite%
{dakna}. But due to severe limitation imposed by coherence loss and damping,
we will employ the scheme introduced by Dakna et al. \cite{dakna} in the
realm of running wave field. For brevity, the present application only shows
the relevant steps of Ref.\cite{dakna}, where the reader will find more
details. In this scheme, a desired state $\left\vert \Psi \right\rangle $
composed of a finite number of Fock states $\left\vert n\right\rangle $ can
be written as%
\begin{eqnarray}
\left\vert \Psi \right\rangle &=&\sum_{n=0}^{N}C_{n}\left\vert
n\right\rangle =\frac{C_{N}}{\sqrt{N!}}\prod_{n=1}^{N}\left( \hat{a}%
^{+}-\beta _{n}^{\ast }\right) \left\vert 0\right\rangle  \nonumber \\
&=&\frac{C_{N}}{\sqrt{N!}}\prod_{k=1}^{N}\hat{D}(\beta _{k})\hat{a}^{+}\hat{D%
}(\beta _{k})\left\vert 0\right\rangle ,  \label{teorica}
\end{eqnarray}%
where $\hat{D}(\beta _{n})$ stands for the displacement operator and the $%
\beta _{n}$ are the roots of the polynomial equation
\begin{equation}
\sum\limits_{n=0}^{N}C_{n}\beta ^{n}=0.  \label{raizes}
\end{equation}

According to the experimental setup shown in the Fig.1 of Ref.\cite{dakna},
we have (assuming $0$-photon registered in all detectors) that the outcome
state is%
\begin{equation}
\left\vert \Psi \right\rangle \sim \prod_{k=1}^{N}D(\alpha _{k+1})\hat{a}%
^{+}T^{\hat{n}}D(\alpha _{k}) \left\vert 0\right\rangle ,
\label{experimental}
\end{equation}%
where $T$ is the transmittance of the beam splitter and ${\alpha }_{k}$ are
experimental parameters. After some algebra, the Eq.(\ref{teorica}) and Eq.(%
\ref{experimental}) can be connected. In this way, one shows that they
become identical when ${\alpha }_{1}=-\sum_{l=1}^{N}T^{-l}{\alpha }_{l+1}$
and $\alpha _{k}=T^{\ast N-k+1}(\beta _{k-1}-\beta _{k})$ for $k=2,3,4\ldots
N$. \ In the present case the coefficients $C_{n}$ are given by those of the
TSI. The roots $\beta _{k}^{\ast }=\left\vert \beta _{k}\right\vert
e^{-i\varphi _{\beta _{k}}}$ of the characteristic polynomial in Eq.(\ref%
{raizes}) and the displacement parameters $\alpha _{k}=\left\vert \alpha
_{k}\right\vert e^{-i\varphi _{\alpha _{k}}}$ are shown in the Tables
I;II;III and IV, for $N=5.$

%%%%%%%%%%%%%%%%%%%%%%%%%%%%%%%%%%%%%%%%%%%%%%%%%%%%%%%%%%%%%%
\begin{table}[tbp]
\caption{The roots $\protect\beta _{k}^{\ast }=|\protect\beta _{k}|e^{-i%
\protect\varphi _{\protect\beta _{k}}}$ of the characteristic polynomial and
the displacement parameters $\protect\alpha _{k}^{\ast }=|\protect\alpha %
_{k}|e^{-i\protect\varphi _{\protect\alpha _{k}}}$ are given for TSI using
the doubling function for $C_{0}=0.3$ (coinciding with nonchaotic behavior
in the DST sense), $N=5$ and $T=0.862$. The probability of producing the
state is 0.22$\%.$}\centering$%
\begin{tabular}{||c||c||c||c||c||}
\hline\hline
N & $\left| \beta _{k}\right| $ & $\varphi _{\beta _{k}}$ & $\left| \alpha
_{k}\right| $ & $\varphi _{\alpha _{k}}$ \\ \hline\hline
1 & 2.169 & 2.638 & 1.187 & -0.220 \\ \hline\hline
2 & 2.169 & -2.638 & 1.155 & 1.570 \\ \hline\hline
3 & 0.545 & 3.141 & 1.096 & -2.483 \\ \hline\hline
4 & 1.460 & 1.084 & 1.323 & -2.331 \\ \hline\hline
5 & 1.460 & -1.084 & 2.225 & 1.570 \\ \hline\hline
6 &  &  & 1.460 & -1.084 \\ \hline\hline
\end{tabular}%
$%
\end{table}
%
%%%%%%%%%%%%%%%%%%%%%%%%%%%%%%%%%%%%%%%%%%%%%%%%%%%%%%%%%

%%%%%%%%%%%%%%%%%%%%%%%%%%%%%%%%%%%%%%%%%%%%%%%%%%%%%%%%%%%%%%
\begin{table}[tbp]
\caption{The roots $\protect\beta _{k}^{\ast }=|\protect\beta _{k}|e^{-i%
\protect\varphi _{\protect\beta _{k}}}$ of the characteristic polynomial and
the displacement parameters $\protect\alpha _{k}^{\ast }=|\protect\alpha %
_{k}|e^{-i\protect\varphi _{\protect\alpha _{k}}}$ are given for a TSI using
the doubling function for $C_{0}=0.29711$ (coinciding with chaotic behavior
in the DST sense), $N=5$ and $T=0.867$. The probability of producing the
state is 0.21$\%$}\centering$%
\begin{tabular}{||c||c||c||c||c||}
\hline\hline
N & $\left| \beta _{k}\right| $ & $\varphi _{\beta _{k}}$ & $\left| \alpha
_{k}\right| $ & $\varphi _{\alpha _{k}}$ \\ \hline\hline
1 & 2.306 & 2.692 & 1.372 & -0.198 \\ \hline\hline
2 & 2.306 & -2.692 & 1.130 & 1.570 \\ \hline\hline
3 & 0.543 & 3.141 & 1.193 & -2.563 \\ \hline\hline
4 & 1.489 & 1.089 & 1.357 & -2.321 \\ \hline\hline
5 & 1.489 & -1.089 & 2.289 & 1.570 \\ \hline\hline
6 &  &  & 1.489 & -1.089 \\ \hline\hline
\end{tabular}%
$%
\end{table}
%
%%%%%%%%%%%%%%%%%%%%%%%%%%%%%%%%%%%%%%%%%%%%%%%%%%%%%%%%%

%%%%%%%%%%%%%%%%%%%%%%%%%%%%%%%%%%%%%%%%%%%%%%%%%%%%%%%%%%%%%%
\begin{table}[t]
\caption{The roots $\protect\beta _{k}^{\ast }=|\protect\beta _{k}|e^{-i%
\protect\varphi _{\protect\beta _{k}}}$ of the characteristic polynomial and
the displacement parameters $\protect\alpha _{k}^{\ast }=|\protect\alpha %
_{k}|e^{-i\protect\varphi _{\protect\alpha _{k}}}$ are given for TSI using
the logistic function for $C_{0}=0.2$, $\protect\mu =3.49$ (coinciding with
nonchaotic behavior in the DST sense), $N=5$ and $T=0.893$. The probability
of producing the state is 0.11$\%$}\centering$%
\begin{tabular}{||c||c||c||c||c||}
\hline\hline
N & $\left| \beta _{k}\right| $ & $\varphi _{\beta _{k}}$ & $\left| \alpha
_{k}\right| $ & $\varphi _{\alpha _{k}}$ \\ \hline\hline
1 & 3.948 & 3.141 & 2.794 & 0.051 \\ \hline\hline
2 & 0.609 & 2.566 & 2.195 & -3.045 \\ \hline\hline
3 & 0.609 & -2.566 & 0.472 & 1.570 \\ \hline\hline
4 & 1.828 & 1.373 & 1.830 & -1.959 \\ \hline\hline
5 & 1.828 & -1.373 & 3.202 & 1.570 \\ \hline\hline
6 &  &  & 1.828 & -1.373 \\ \hline\hline
\end{tabular}%
$%
\end{table}
%
%%%%%%%%%%%%%%%%%%%%%%%%%%%%%%%%%%%%%%%%%%%%%%%%%%%%%%%%%

%%%%%%%%%%%%%%%%%%%%%%%%%%%%%%%%%%%%%%%%%%%%%%%%%%%%%%%%%%%%%%
\begin{table}[t]
\caption{The roots $\protect\beta _{k}^{\ast }=|\protect\beta _{k}|e^{-i%
\protect\varphi _{\protect\beta _{k}}}$ of the characteristic polynomial and
the displacement parameters $\protect\alpha _{k}^{\ast }=|\protect\alpha %
_{k}|e^{-i\protect\varphi _{\protect\alpha _{k}}}$ are given for TSI using
the logistic function for $C_{0}=0.2,$ $\protect\mu =4$ (coinciding with
chaotic behavior in the DST sense), $N=5$ and $T=0.879$. The probability of
producing the state is 0.15$\%$}\centering$%
\begin{tabular}{||c||c||c||c||c||}
\hline\hline
N & $\left| \beta _{k}\right| $ & $\varphi _{\beta _{k}}$ & $\left| \alpha
_{k}\right| $ & $\varphi _{\alpha _{k}}$ \\ \hline\hline
1 & 3.290 & 3.141 & 2.027 & 0.094 \\ \hline\hline
2 & 0.563 & 2.708 & 1.665 & -3.056 \\ \hline\hline
3 & 0.563 & -2.708 & 0.321 & 1.570 \\ \hline\hline
4 & 1.893 & 1.255 & 1.787 & -2.064 \\ \hline\hline
5 & 1.893 & -1.255 & 3.165 & 1.570 \\ \hline\hline
6 &  &  & 1.893 & -1.255 \\ \hline\hline
\end{tabular}%
$%
\end{table}
%
%%%%%%%%%%%%%%%%%%%%%%%%%%%%%%%%%%%%%%%%%%%%%%%%%%%%%%%%%

For $N=5$, the best probability of producing TSI is $0.22\%$ when the
doubling function is used, and $\ 0.15\%$ when the logistic funcion is used.
The beam-splitter transmittance which optimizes this probability is around $%
T=0.878$.

\section{Fidelity of generation of TSI}

Until now we have assumed all detectors and beam-splitters as ideal.
Although very good beam-splitters are available by advanced technology, the
same is not true for photo-detectors in the optical domain. Thus, let us now
take into account the quantum efficiency $\eta $ at the photodetectors. For
this purpose, we use the Langevin operator technique as introduced in \cite%
{norton1} to obtain the fidelity to get the TSI.

Output operators accounting for the detection of a given field $\hat{\alpha}$
reaching the detectors are given by \cite{norton1}

\begin{equation}
\widehat{\alpha }_{out}=\sqrt{\eta }\widehat{\alpha }_{in}+\widehat{L}%
_{\alpha },  \label{E9}
\end{equation}%
where $\eta $ stands for the efficiency of the detector and $\widehat{L}%
_{\alpha }$, acting on the environment states, is the noise or Langevin
operator associated with losses into the detectors placed in the path of
modes $\widehat{\alpha }=a,b$. We assume that the detectors couple neither
different modes $a,b$ nor the Langevin operators $\widehat{L}_{\alpha }$, so
the following commutation relations are readily obtained from Eq.(\ref{E9}):

%\begin{mathletters}
\begin{eqnarray}
\left[ \widehat{L}_{\alpha },\widehat{L}_{\alpha }^{\dagger }\right]
&=&1-\eta ,  \label{E10a} \\
\left[ \widehat{L}_{\alpha },\widehat{L}_{\beta }^{\dagger }\right] &=&0.
\label{E10b}
\end{eqnarray}%
The ground-state expectation values for pairs of Langevin operators are

%\end{mathletters}
%\begin{mathletters}
\begin{eqnarray}
\left\langle \widehat{L}_{\alpha }\widehat{L}_{\alpha }^{\dagger
}\right\rangle &=&1-\eta ,  \label{E11a} \\
\left\langle \widehat{L}_{\alpha }\widehat{L}_{\beta }^{\dagger
}\right\rangle &=&0,  \label{E11b}
\end{eqnarray}%
which are useful relations specially for optical frequencies, when the state
of the environment can be very well approximated by the vacuum state, even
for room temperature.

Let us now apply the scheme of the Ref.\cite{dakna} to the present case. For
simplicity we will assume all detectors having high efficiency ($\eta
\gtrsim 0.9$). This assumption allows us to simplify the resulting
expression by neglecting terms of order higher than $(1-\eta )^{2}$. When we
do that, instead of $|TSI\rangle $, we find the (mixed) state $|\Psi
_{FE}\rangle $ describing the field plus environment, the latter being due
to losses coming from the nonunit efficiency detectors. We have,
%\end{mathletters}
\begin{eqnarray}
\left\vert \Psi _{FE}\right\rangle &\sim &R^{N}D(\alpha _{N+1})\hat{a}%
^{\dagger }T^{\hat{n}}D(\alpha _{N})\hat{a}^{\dagger }T^{\hat{n}}  \nonumber
\\
&\times &D(\alpha _{N-1})\ldots \hat{a}^{\dagger }T^{\hat{n}}D(\alpha
_{1})\left\vert 0\right\rangle \widehat{L}_{0}^{\dagger }  \nonumber \\
&+&R^{N-1}D(\alpha _{N+1})\hat{a}^{\dagger }T^{\hat{n}}D(\alpha _{N})\hat{a}%
^{\dagger }T^{\hat{n}}  \nonumber \\
&\times &D(\alpha _{N-1})\ldots \widehat{L}_{1}^{\dagger }T^{\hat{n}%
}D(\alpha _{1})\left\vert 0\right\rangle  \nonumber \\
&+&R^{N-1}D(\alpha _{N+1})\hat{a}^{\dagger }T^{\hat{n}}D(\alpha _{N})%
\widehat{L}_{N-1}^{\dagger }T^{\hat{n}}  \nonumber \\
&\times &D(\alpha _{N-1})\ldots \hat{a}^{\dagger }T^{\hat{n}}D(\alpha _{1})
\left\vert 0\right\rangle  \nonumber \\
&+&R^{N-1}D(\alpha _{N+1})\widehat{L}_{N}^{\dagger }T^{\hat{n}}D(\alpha _{N})%
\hat{a}^{\dagger }T^{\hat{n}}  \nonumber \\
&\times &D(\alpha _{N-1})\ldots \hat{a}^{\dagger }T^{\hat{n}}D(\alpha _{1})
\left\vert 0\right\rangle ,  \label{damped}
\end{eqnarray}%
where, for brevity, we have omitted the kets corresponding to the
environment. Here $R$ is the reflectance of the beam splitter, $\ \widehat{L}%
_{0}^{\dagger }=\mathbf{1}$ is the identity operator and $\widehat{L}_{k}$, $%
k=1,2..N$ stands for losses in the first, second $\ldots $ $N-th$ detector.
Although the $\widehat{L}_{k}\prime s$ commute with any system operator, we
have maintained the order above to keep clear the set of possibilities for
photo absorption: the first term, which includes $\widehat{L}_{0}^{\dagger }=%
\mathbf{1}$, indicates the probability for nonabsorption; the second term,
which include $\widehat{\mathsf{L}}_{1}^{+}$, indicates the probability for
absorption in the first detector; and so on. Note that in case of absorption
at the k-$th$ detector, the annihilation operator $a$ is replaced by the $%
\widehat{L}_{k}^{\dagger }$ creation Langevin operator. Other possibilities
such as absorption in more than one detector lead to a probability of order
lesser than $(1-\eta )^{2}$, which will be neglected.

Next, we have to compute the fidelity \cite{nota}, $F=\left\Vert
\left\langle \Psi \right. \left\vert \Psi _{FE}\right\rangle \right\Vert
^{2} $, where $\left\vert \Psi \right\rangle $ is the ideal state given by
Eq.(\ref{experimental}), here corresponding to the TSI characterized by the
parameters shown in Tables I-IV, and $\left\vert \Psi _{FE}\right\rangle $
is the state given in the Eq.(\ref{damped}). Assuming $\eta =$ $0.99$, $0.95$
and $0.90$ and starting with $C_{0}=0.3$ and $C_{0}=0.29711$ for the
doubling function, we find $F\simeq 0.9983$, $0.9943$ and $0.9909$,
respectively, and for the logistic function, starting with $C_{0}=0.2$, $\mu
=3.49$ and $\mu =4$, we find $F\simeq 0.9986$, $0.9944$ and $0.9911$,
respectively. These high fidelities show that efficiencies around $0.9$ lead
to states whose degradation due to losses is not so dramatic for $N=5$.

\section{Comments and conclusion}

In this paper we have introduced new states of the quantized electromagnetic
field, named truncated states with probability amplitudes obtained through
iteration of a function (TSI). Although TSI can be building using various
functions such as logistic, sine, exponential functions and so on, we have
focused our attention on the doubling and the logistic functions, which, as
is well known from dynamical systems theory, can exhibit a chaotic behavior
in the interval (0,1]. To characterize the TSI for the doubling and logistic
functions we have studied various of its features, including some
statistical properties, as well as the behavior of these features when the
dimension $N$ of Hilbert space is increased. Interesting, we found a
transition from sub-poissonian statistics to super-poissonian statistics
when $N$ is relatively small ($N\sim 12$). Besides, photon number
distribution, which is analogous to concept of orbits in the study of the
dynamic of maps, shows a regular or rather a \textquotedblleft
chaotic\textquotedblright\ behavior depending on existing or not fixed or
periodic points in the function to be iterated. Interestingly enough, we
have found a pattern when the properties of \ TSI for logistic function are
compared with that of TSI for the doubling function from the point of view
of dynamical systems theory (DST). For example, as $P_{n}$ has an analog
with orbits from DST, it is straightforward to identify repetitions (or
\textit{periods}) in $P_{n}$, if there are any, when the Hilbert space is
increased. Surprisingly, although the doubling and the logistic function are
different from each other, when other properties such as even and odd photon
number distribution, the average number and its variance, the Mandel
parameter and the second order correlation function were studied, they
presented the same following pattern: if, from the point of view of DST, the
coefficients of TSI for the doubling and the logistic function correspond to
a nonchaotic (chaotic) regime, all those properties increases smoothly
(irregularly) when the Hilbert space is increased.

\section{\textbf{Acknowledgments}}

NGA thanks CNPq, Brazilian agency, and VPG-Universidade Cat\'{o}lica de Goi%
\'{a}s, and WBC thanks CAPES, for partially supporting this work.

\end{document}